\documentclass[epjST]{svjour}
\usepackage{graphics}
\usepackage{color}
\usepackage{amssymb}

\begin{document}
\title{Momentum distribution of Vinen turbulence in trapped atomic Bose-Einstein condensates}
\author{ \'Attis V. M. Marino\inst{1} \and Lucas Madeira\inst{1}\fnmsep\thanks{\email{madeira@ifsc.usp.br}} \and Andr\'e Cidrim\inst{2} \and F. E. A. dos Santos\inst{2} \and Vanderlei S. Bagnato\inst{1,3}}
\institute{Instituto de F\'isica de S\~ao Carlos, Universidade de S\~ao Paulo, CP 369, S\~ao Carlos, S\~ao Paulo 13560-970, Brazil \and Departamento de F\'isica, Universidade Federal de S\~ao Carlos, S\~ao Carlos, 13565-905, Brazil \and
Hagler Fellow, Department of Biomedical Engineering, Texas A\&M University, College Station, Texas 77843, USA
}
\abstract{
The decay of multicharged vortices in trapped Bose-Einstein condensates may lead to a disordered vortex state consistent with the Vinen regime of turbulence, characterized by an absence of large-scale flow and an incompressible kinetic energy spectrum $E\propto k^{-1}$. In this work, we study numerically the dynamics of a three-dimensional harmonically trapped Bose-Einstein condensate excited to a Vinen regime of turbulence through the decay of two doubly-charged vortices. First, we study the momentum distribution and observe the emergence of a power-law behavior $n(k)\propto k^{-3}$ consistent with the coexistence of wave turbulence. We also study the kinetic energy and particle fluxes, which allows us to identify a direct particle cascade associated with the turbulent stage.
} 
\maketitle

\textit{Introduction:}
The field of quantum turbulence deals with the manifestation
of turbulence in quantum fluids \cite{tsatos16,madeira20}.
Although there is a small range of length scales available in trapped Bose-Einstein condensates (BECs), much progress has been made in studying turbulence in these systems.
Under appropriate conditions, statistical properties
of classical turbulence may arise, such as the
Kolmogorov scaling of the energy spectrum ($E\propto k^{-5/3}$).
Experiments \cite{walmsley08} showed that,
besides Kolmogorov turbulence, there is another
regime where superfluid $^4$He displays a different kinetic energy
spectrum. This other type of turbulence, called Vinen or
ultraquantum turbulence, was observed in superfluid helium by
controlling the injections of vortex rings in the system.
A diagnostic to differentiate both kinds of turbulence is the
temporal decay of the vortex line density $L(t)$.
Numerical simulations of the experiment \cite{baggaley12,baggaley14}
showed that Vinen turbulence,
corresponding to $L(t)\propto t^{-1}$, has a spectrum $E\propto k^{-1}$
for large $k$ (in the hydrodynamical range), while the $-5/3$ Kolmogorov
scaling was observed for the $L(t)\propto t^{-3/2}$ regime.
In Ref.~\cite{cidrim17}, a procedure to generate turbulence in a trapped
Bose-Einstein condensate through the decay of multicharged vortices
was proposed, where the authors associated the resultant disordered vortex state with the Vinen regime of turbulence.
In this paper, we use their results to explore the momentum distribution $n(k)$ in this regime. We found a power-law behavior $n(k)\propto k^{-3}$ that is consistent with a wave-turbulent regime with direct particle cascade, provided by an energy and particle flux analysis.

\textit{Particle and energy fluxes:}
The energy and particle fluxes have been successfully used to further characterize quantum turbulence in numerical simulations of bulk helium \cite{baggaley14} and homogeneous condensates \cite{muller20}, and also in experiments with box-trapped \cite{navon19} and harmonically confined BECs \cite{orozco20}. Hence, computing these quantities in numerical simulations of the Vinen regime of turbulence in trapped BECs may also bring insight into aspects that might not be directly probed by experiments.
The total number of particles is conserved in the numerical simulations of the Gross-Pitaevskii equation (GPE) employed
in this work.
To quantify the particle transfer from large to small length scales, we computed the
particle flux,
\begin{equation}
\label{eq:PiN}
\Pi_N(k)=-\int_{k_0}^{k} \frac{\partial n(k')}{\partial t} dk',
\end{equation}
where $k_0$ is inversely proportional to the largest length scale of the system. We adopted a sign convention where $\Pi_N(k)>0$ means that
particles are being transferred from large to small length scales, and in the opposite
direction for a negative particle flux.
Equation~(\ref{eq:PiN}) is simply a consequence of the continuity equation, and its derivation can be found in Refs.~\cite{navon19,orozco20}.
The total energy is also a conserved quantity in these simulations. However, we are interested mainly in the kinetic energy component, which has been used as a diagnostic tool to identify the turbulent regime. 
To quantify the kinetic energy transfer from large to small length scales, we computed the
energy flux
\begin{equation}
\label{eq:PiE}
\Pi_E(k)=-\int_{k_0}^{k} \frac{\partial E(k')}{\partial t} dk',
\end{equation}
where the conventions are the same as the ones employed in Eq.~(\ref{eq:PiN}). Analogously, this is also a consequence of the continuity equation.

\textit{Decay of two doubly charged vortices in a harmonic trap:}
Topological phase imprint has been used to generate
multicharged vortices \cite{shin04} in atomic BECs. In particular,
the decay of a doubly quantized vortex into two singly charged ones has
been studied in Refs.~\cite{shin04,huhtamaki06,mateo06}.
In this work, we employ the simulations of Ref.~\cite{cidrim17}.
The condensate's dynamics is given by the 3D GPE,
\begin{equation}
i\hbar\frac{\partial \Psi}{\partial t}=
\left(
-\frac{\hbar^2}{2m}\nabla^2+V(\textbf{r})+g|\Psi|^2,
\right)\Psi
\end{equation}
where $\Psi(\textbf{r},t)$ is the macroscopic wave function at the position
$\textbf{r}$ and time $t$, and the trapping potential
$V(\textbf{r})=m(\omega_r^2 r^2+\omega_z^2 z^2)/2$,
where $r^2=x^2+y^2$. The parameter $g=4\pi\hbar^2a_s/m$, where $a_s$ is the
s-wave scattering length, was chosen to be $g=8600$, to reproduce a typical
$^{87}$Rb condensate, and the radial and axial frequencies such that $\omega_z/\omega_r=0.129$.
We work with dimensionless quantities by using $\tau_{HO}=\omega_r^{-1}$,
$\ell_{HO}=\sqrt{\hbar/m\omega_r}$, $\hbar\omega_r$ as units of time, length, and energy, respectively.
The transverse and axial Thomas-Fermi radii are $d_{TF}=4.2 \ell_{HO}$ and
$D_{TF}=32.6\ell_{HO}$.

The momentum distribution is related to the Fourier transform of the wave function, $n(k)=|\tilde{\psi}(k)|^2$.
For the kinetic energy spectrum, first we compute the velocity field in real space, $\textbf{v}(\textbf{r})$, through the continuity equation.
From it, we can define a density-weighted velocity field
$\textbf{w}(\textbf{r})=\sqrt{|\psi(\textbf{r})|^2}\textbf{v}(\textbf{r})$.
Using the Helmholtz theorem we can separate the compressible and incompressible components,
$\textbf{w}(\textbf{r})=\textbf{w}_c(\textbf{r})+\textbf{w}_i(\textbf{r})$.
Finally, we can compute the Fourier transform of $\textbf{w}_i(\textbf{r})$ to calculate the spectrum
$E(k)=(m/2)|\tilde{\textbf{w}}_i(k)|^2$.

\begin{figure}[!htb]
\centering
\resizebox{0.95\columnwidth}{!}{
\includegraphics{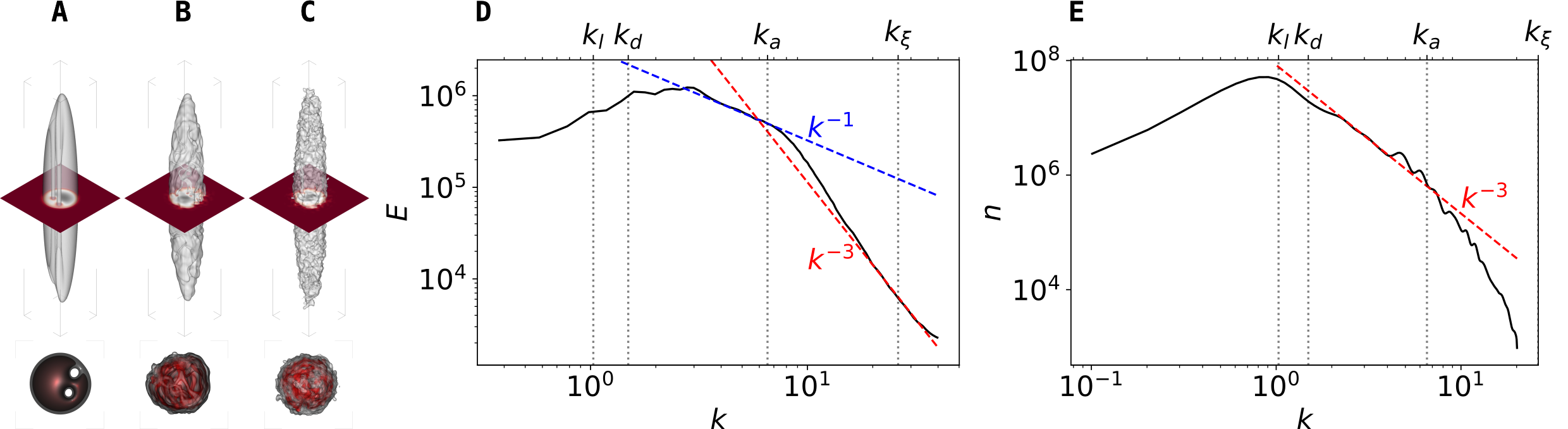}
}
\caption{
Isodensity plots (a-c) of the cloud and a cross section, showing the time evolution of two initial doubly-charged antiparallel vortices (a) into the turbulent quasi-isotropic state at $t=12.8 \tau_{\rm HO}$(b), which has finally decayed at long times, $t=50.0 \tau_{\rm HO}$ (c).
Incompressible kinetic energy spectrum for $t=12.8\tau_{\rm HO}$ (d). It is possible to see a region where $E\propto k^{-1}$, consistent with the Vinen regime of turbulence, and another where $E\propto k^{-3}$, result of the vortex core structure.
Momentum distribution of the turbulent cloud at the same instant (e). Note the appearance of a power-law behavior $n(k) \propto k^{-3}$ for $k_\ell\lesssim k \lesssim k_a$. The absence of self-similar vortex dynamics in this $k$-range corroborates the occurrence of a simultaneous wave-turbulence regime.}
\label{fig:spectra}
\end{figure}

\textit{Results and Discussion:}
The initial system is prepared by imprinting two doubly charged antiparallel
vortices in a cigar-shaped BEC.
We computed the incompressible kinetic energy spectrum, which is shown in Fig.~\ref{fig:spectra}a for $t=12.8 \tau_{\rm HO}$.
The wavenumbers corresponding to characteristic length scales of the system are indicated in the figure, $k_D=2\pi/D_{TF}$, $k_\ell=2\pi/\ell$, $k_d=2\pi/d_{TF}$, $k_a=2\pi/a$, $k_\xi=2\pi/\xi$, with $a=0.96\ell_{HO}$ being the vortex core size and $\xi=0.24\ell_{HO}$ the healing length.
Clearly, the energy spectrum is not Kolmogorov's, and the scaling is proportional
to $k^{-1}$, which is characteristic of a single isolated straight vortex line. So, for distances smaller than $\ell$, the nearest vortex to the
observation point dominates. The $k^{-3}$ scaling, around $k_\xi$ is characteristic of a vortex core \cite{bradley12}. Lastly, the authors observed the temporal decay of the vortex line density, which was proportional to $t^{-1}$. Putting together all these pieces of evidence, it is possible to identify this state as in the Vinen turbulent regime.

It is known that, alongside the chaotic dynamics of quantized vortices, there can be nonlinear interactions of waves, which create yet another type of turbulence in BECs~\cite{nazarenko11}. The theory behind this wave turbulence has successfully described the limit of weak-wave interactions. It offers well-defined predictions for the steady-state momentum distribution $n(k)$ of a perturbed BEC. With this in mind, we also computed the momentum distribution for our Vinen turbulent cloud and observed the emergence of a power-law, with $n(k) \propto k^{-3}$ for $k_\ell\lesssim k \lesssim k_a$ (see Fig.~\ref{fig:spectra}b). The presence of strong coherent structures, i.e., quantized vortices, and the anisotropy imposed by the bounds of a trapped condensate offer however great theoretical challenges. No analytical predictions are thus available for this system. Despite that, results from weak-wave turbulence theory have predicted similar power-laws, $n(k)\propto k^{-3}$~\cite{fujimoto15} and $n(k)\propto k^{-7/2}$~\cite{nazarenko11}, for homogeneous BECs. Our calculation is also consistent with the $n(k)\sim k^{-2.9}$ found in turbulence experiments using the same trap parameters we use in our simulations~\cite{thompson13} and close to analogous measurements in box-trapped turbulent BECs~\cite{navon16}. Therefore, the absence of self-similar vortex dynamics (a particularity of Vinen turbulence) in the same $k$-range where we observe the power-law behavior in $n(k)$ suggests a coexisting wave-turbulent regime. This can be understood as a result of nonlinear interaction of waves produced from vortex reconnections, which can, in principle, be density fluctuations or even Kelvin waves~\cite{nazarenko06}.

\vspace{-0.5cm}

\vspace{-0.5cm}

\begin{figure}[!htb]
\centering
\resizebox{0.65\columnwidth}{!}{
\includegraphics{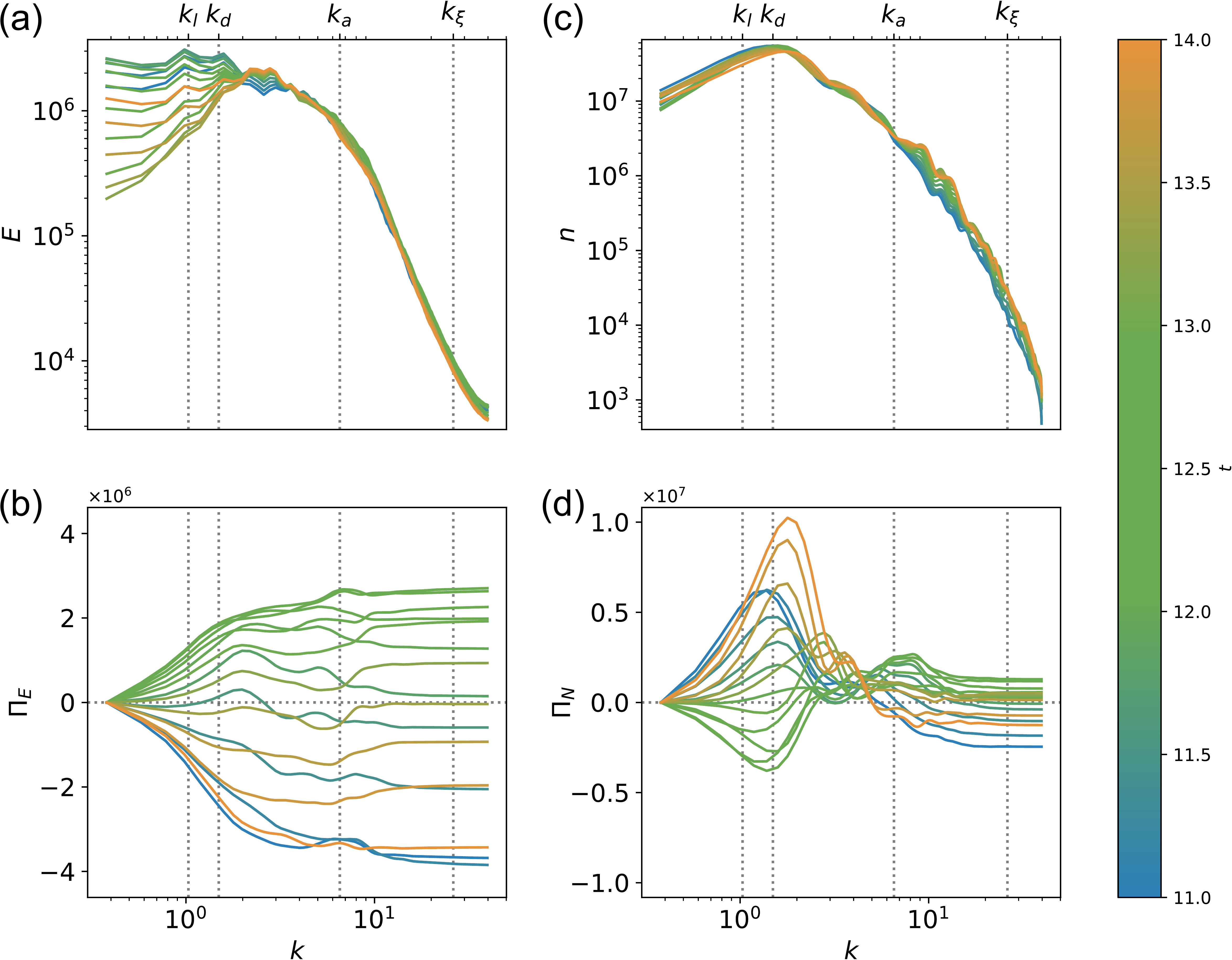}}
\caption{Kinetic energy spectrum (a) and flux (b), momentum distribution (c) and particle flux (d) as a function of the wave number $k$ for $11\leqslant t \leqslant 14$. Both fluxes present an oscillatory behavior with frequency 2.0(1)$\omega_r$.}
\label{fig:fluxes}
\end{figure}

Using Eqs.~(\ref{eq:PiN}) and (\ref{eq:PiE}) we calculated the kinetic energy and particle fluxes during the
time evolution of the system.
We found that both fluxes present an oscillatory behavior, with a characteristic angular frequency of 2.0(1)$\omega_r$, which can be observed for all times in the simulation.
This is consistent with a breathing mode, a feature also found in Ref.~\cite{cidrim17} using a different method. Since we are interested in the interval where Vinen turbulent regime is observed, in Fig.~\ref{fig:fluxes} we present the results for $11\leqslant t \leqslant 14$.

\vspace{-0.55cm}

\vspace{-0.65cm}

The kinetic energy flux could, in principle, be used to identify the direction of the energy cascade. It has recently been applied to understand the effects of dimensionality transition in turbulent homogeneous condensates \cite{muller20}. One may be tempted to analogously infer the direction of kinetic energy transfer in our turbulent trapped system. This quantity however is highly oscillatory (see Fig.~\ref{fig:fluxes}a), as a result of finite-size effects, and suggests exchange with other energy components (particularly potential energy). This is corroborated by abrupt changes in the sign at the frequency of the collective breathing mode. The kinetic energy is not the conserved quantity in the system (the actual energy conservation in the GPE regards the total energy) and implies that an analysis to determine the direction of kinetic energy transport in our trapped system through its flux (as defined in Eq.~\ref{eq:PiE}) becomes inconclusive.

This limitation nevertheless is not present for the particle flux, since $dN/dt=0$ throughout the entire evolution. We found that the largest variations of the particle flux occur at low values of $k$, close to $k_d$, corresponding to the transverse Thomas-Fermi radius. Again, this feature is consistent with the collective breathing mode of the cloud at large length scales ($k\lesssim k_d$), which should not interfere much with the $k$-region where the Vinen scaling is observed, $3.8\leqslant k \leqslant 6.4$. For the time interval where we identify the statistical properties of Vinen turbulence, we note a change in particle flux, which becomes positive and almost constant for the $k$-range where the incompressible kinetic energy displays the $k^{-1}$ scaling, up to $k \sim k_\xi$. This indicates a direct particle cascade on top of the chaotic tangling of vortices in Vinen turbulence. A peak in the flux can also be identified at $k\approx k_a$, a probable signature of vortex reconnections at scales of the order of a vortex core.

In conclusion, we calculated the momentum distribution of a trapped BEC in the Vinen turbulent regime. We identified the coexistence of wave turbulence for scales where the incompressible kinetic energy displays the $k^{-1}$ power law. We also computed the kinetic energy and particle fluxes, with both exhibiting an oscillatory behavior at the same frequency as the collective breathing mode of the trapped cloud. While finding the direction of kinetic energy transport through analyzing its flux is prevented by finite-size limitations, the analysis of particle flux allows us to observe a direct particle cascade, coinciding with a wave-turbulent energy cascade resulting from vortex reconnection events. The nature of such a cascade, whether originated from interactions of density or Kelvin waves, is an exciting topic for future investigation.

\vspace{-0.3cm}

\begin{acknowledgement}
This work was supported by
the S\~ao Paulo Research Foundation (FAPESP)
under the grants 2013/07276-1, 2014/50857-8, 2017/09390-7, and 2018/09191-7, and by the
National Council for Scientific and Technological Development (CNPq)
under the grant 465360/2014-9. 

Author contributions:
conceptualization, F.E.A.S and V.S.B;
methodology, L.M. and A.C.;
software, A.V.M.M and A.C.;
formal analysis, L.M. and A.C.;
writing--original draft preparation, L.M. and A.C.;
writing--review and editing, F.E.A.S and V.S.B;
visualization, A.V.M.M.
All authors have read and agreed to the published version of the manuscript.
\end{acknowledgement}

\vspace{-0.65cm}

\bibliographystyle{unsrt}
\bibliography{article}

\end{document}